\documentclass{elsart3p_tech_mod}

\usepackage{amsfonts,latexsym,citesort,amssymb,amsmath}
\usepackage{url}
\usepackage{graphicx,subfigure}
\usepackage{capt-of}
\usepackage[mathscr]{eucal}
\usepackage{comment}
\usepackage{color}

\newcommand{\bits}{{\{0,1\}}}

\def\mod{{\mbox{~\textrm{mod}~}}}

\begin{document}

\begin{frontmatter}

\title{Attacking the Diebold Signature Variant -- RSA Signatures with Unverified High-order Padding}

\author[jhu]{Ryan W. Gardner\thanksref{cor}},
\corauth[cor]{Corresponding author.}
\ead{ryan@cs.jhu.edu}
\author[washu]{Tadayoshi Kohno},
\ead{yoshi@cs.washington.edu}
\author[fsu]{Alec Yasinsac}
\ead{yasinsac@cs.fsu.edu}

\address[jhu]{Department of Computer Science, Johns Hopkins University, Baltimore, MD 21218, USA}
\address[washu]{Department of Computer Science and Engineering, University of Washington, Seattle, WA 98195, USA}
\address[fsu]{Department of Computer Science, Florida State University, Tallahassee, FL 32306, USA}

\begin{abstract}
We examine a natural but improper implementation of RSA signature
verification deployed on the widely used Diebold Touch
Screen and Optical Scan voting machines.  In the implemented scheme,
the verifier fails to examine a large number of the high-order bits of
signature padding and the public exponent is three.  We present an
very mathematically simple attack that enables an adversary to
forge signatures on arbitrary messages in a negligible amount of time.

\end{abstract}


\end{frontmatter}


\section{Background}

Standard signatures using the RSA primitive are generally computed in two
steps: 1) A one-way transformation $T$ is applied to a message $m$ to
produce an encoded message $M$.  
2) The RSA primitive using the private exponent is applied to $M$ to
generate the signature $\sigma$.  Given a signature $\sigma'$ on a
message $m'$, verification generally proceeds similarly: 1) Again, the
transformation $T$ is applied to $m'$ to obtain the encoded message
$M'=T(m')$, and 2) the RSA primitive using the public exponent is applied
to the signature $\sigma'$ to produce $M''$.  3) Finally, $M'$ and
$M''$ are checked for equality, and the signature verifies if and only
if they are equivalent.

In this paper, we examine an improperly implemented RSA signature
scheme, which uses a public exponent of three, where the verifier
fails to compare a large number of the high-order bits of the encoded
messages $M'$ and $M''$.  We briefly present an simple attack against
such implementations that enables an adversary to forge signatures on
arbitrary messages in a negligible amount of time.  Specifically, the
attack works for any transformation function $T$ and for all messages
when $b < \frac{1}{3}\ell_n - 3$, where $b$ is the number of low-order
bits of $M'$ and $M''$ that are examined, and $\ell_n$ is the
bit-length of the RSA modulus.

This flawed signature scheme has become relevant in practice as we
observe multiple instantiations of it in recent (May 2007) versions of
the widely used Diebold voting machine firmware.\footnote{The code was
examined under a charter from the Florida Department of State in June
2007 \cite{florida_report}.}  We find that the Diebold Touch Screen
bootloader 1.3.6 and Optical Scan 1.96.8 employ a natural,
yet flawed, implementation of ``text-book'' RSA where a transformation
function $T=\textsf{SHA-1}$ is used, and the least significant 160
bits (i.e. the \textsf{SHA-1} hash) of $M'$ and $M''$ are exclusively
examined when verifying their equality.  The verified signatures cover
data \cite{florida_report} that, if unauthenticated, has been
publicly reported to enable simple arbitrary software installation
\cite{hursti2}, vote pre-loading (and pre-removing) \cite{hursti1},
arbitrary code execution \cite{wagner_interpreter}, and a mass
spreading, vote stealing virus \cite{princeton} on the voting
machines.

Several other research papers have analyzed variants and flawed
constructions of RSA authentication systems and their padding and
redundancy schemes.  Recently, Bleichenbacher describes an attack
under conditions similar to those we examine, where a public exponent
of three is used and erroneously implemented signature verification
code fails to appropriately verify most of the \emph{low-order} bits
of a \textsf{PKCS-1} padded message \cite{bleichenbacher}.
Examining another standard, Coron, Naccache, and Stern extend a chosen
plaintext attack on the RSA cryptosystem by Desmedt and Odlyzko
\cite{desmedt} to develop a signature forgery strategy that is
effective on a scheme that uses a padding format differing from
\textsf{ISO 9796-1} by only one bit \cite{coron}.  In addition to
these more common modes for RSA, several studies have discovered
weaknesses, under varying conditions, in a number of schemes that do
not hash messages before signing, but instead apply redundancy bits to
them
\cite{dejonge_attack,girault,misarsky_lll,brier,lenstra,michels}.

\section{Construction}

We now describe the construction of the flawed RSA signature scheme we
examine.

Let $n=pq$ be an $\ell_n$-bit standard RSA modulus equal to the
product of two primes, and let $d$ denote the private exponent such
that $3d \equiv 1 \mod \phi(n)$.  Let $3$ be the public exponent.  We
use the notation $[x]_{:a - b}$ to refer to the value obtained by
writing bits $a - b$ of $x$ as a binary number, where bit $0$ is the
least significant bit of $x$.  Furthermore, in the general scheme we
find, a message $m$ is transformed into an encoded message $M$ using a
transformation $T$ before it is signed.  In practice, the function $T$
usually involves hashing and padding the message $m$ or applying
redundancy to it.  However, for the sake of this analysis, we let $T$
be any function $T: \bits^* \rightarrow \mathbb{Z}_n^*$ (where
$\mathbb{Z}_n^*$ may be approximated as all integers inclusively
between $0$ and $n-1$).

Signature generation on a message $m \in \bits^*$ then consists of the
following two classic computations:
\begin{equation} M=T(m) \end{equation}
\begin{equation} \sigma = M^d \mod n \end{equation}
and the resulting signature is then $\sigma \in \mathbb{Z}_n^*$.

\setcounter{equation}{0}

Verification proceeds similarly given a signature $\sigma'$ and a
message $m'$.  The verifier computes:
\begin{equation} M' = T(m') \end{equation}
\begin{equation} M'' = \sigma'^3 \mod n \label{invert} \end{equation}
Then, although an appropriate algorithm would verify that $M' = M''$,
in the faulty construction we find, the verifier checks
\[ [M']_{:0-(b-1)}=[M'']_{:0-(b-1)} \]
for a $b < \frac{1}{3}\ell_n - 3$.  The signature is considered valid
if and only if the two values are equivalent.

\section{Attack}

Assuming the verification scheme described above, we now describe a
very simple method for forging signatures on arbitrary messages.

Let $m$ be any message and $M = T(m)$.  It follows from the
construction that if we can find a value $\sigma$, such that
\[ \sigma^3 \equiv M + 2^b z \mod n \]
for any integer $z$ where $2^b z < n$, then $\sigma$ is a valid
signature.  (Notice that if $2^b z > n$, its value is likely to alter
the lower $b$ bits of the modular reduction, in which case they will
no longer match those of $M$.)

In the case where $M$ is odd, a solution is almost immediate when we
consider the problem in $\mathbb{Z}_{2^b}^*$.  Since we know the order
of $\mathbb{Z}_{2^b}^* = 2^{b-1}$, we can find the cube root of $M
\mod 2^b$.  First, using Euclid's extended algorithm, we find $r$ such
that $3r \equiv 1 \mod 2^{b-1}$.

\begin{claim} $\sigma \equiv M^r \mod 2^b$ is a valid signature on $m$ when $M$ is odd.
\end{claim}

\begin{pf} $\sigma^3 \equiv M^{3r} \equiv M^{1+y \phi(2^b)} \equiv M \mod 2^b$ by Euler's theorem.  Since $\sigma$ was generated as an element of $\mathbb{Z}_{2^b}^*$, we know it can be written as an integer less than $2^b$.  Thus, $\sigma^3 \leq 2^{3b} \leq 2^{\ell_n - 9} < n$
and $\sigma^3 \equiv M + 2^b z \mod n$ for some $z$ where $2^b z < n$.
\end{pf}

In the case where $M$ is even, the same attack does not apply since $M
\notin \mathbb{Z}_{2^b}^*$.  Instead, we use a slightly modified
approach to obtain a member from that group.

Choose $c$ to be the least integer such that $(2^b c)^3>n$, and let
$r$ be defined as above.

\begin{claim}
$\sigma = 2^b c + \left( (M+n)^r \mod 2^b \right)$ is a valid signature
on $m$ when $M$ is even.
\end{claim}

\begin{pf}
Let $\tau = \left( (M+n)^r \mod 2^b \right)$.  We will first find an
upper bound on $\sigma^3$.  Since $\tau < 2^b$, 
\[ \sigma^3 = \left( 2^b c + \tau \right)^3 < \left( 2^b (c+1) \right)^3.\]
Further, $\left( 2^b (c-1) \right)^3 < n$ by our choice of $c$, which
we can rewrite
\[ c+1 < \frac{ \sqrt[3]{n} }{2^b} + 2. \]
Combining these two inequalities yields
\[  \sigma^3 < \left( 2^b \left(\frac{ \sqrt[3]{n} }{2^b} + 2\right) \right)^3 =\left( \sqrt[3]{n} + 2^{b+1} \right)^3, \]
or, by our bound on $b$,
\[ \sigma^3 < \left( \sqrt[3]{n} + 2^{(\frac{1}{3}\ell_n-3)+1} \right)^3 = \left( \sqrt[3]{n} + 2^{-2}\sqrt[3]{n} \right)^3. \]
Thus,
\[ \sigma^3 < \frac{125}{64} n . \]

Now, with the above bound, and the fact that $\sigma^3 > n$, by our
choice of $c$. We can conclude $(\sigma^3 \mod n) = \sigma^3 - n$.
Hence, $\sigma$ is a successful forgery iff $\sigma^3 - n \equiv M \mod 2^b$.
Indeed,
\begin{align*}
\sigma^3-n &\equiv (2^b c)^3 + 3 (2^b c)^2 \tau + 3 (2^b c) \tau ^2 \\
& \hspace{11mm} + \tau^3 - n \mod 2^b \\ &\equiv \tau^3 - n \mod 2^b.
\end{align*}
Since $n$ is odd, $M+n$ is a member of $\mathbb{Z}_{2^b}^*$, and
$\tau^3 \equiv M + n \mod 2^b$ just as we saw in the $M$ odd case.
Therefore, $\sigma^3 \equiv M \mod n \mod 2^b$.
\end{pf}

\section{Conclusion}

We have found a natural, but improper implementation of an RSA
signature scheme in a highly sensitive application, electronic voting.
We present an extremely mathematically simple attack against it, which
is also very practical.\footnote{We verified the attack for the odd
  case on an optical scan machine provided by the state of Florida.}
We hope this subtle, but real flaw emphasizes the challenges of
properly securing systems and also stresses the importance of
approaching such tasks diligently.\footnote{After being notified of
  the flaw, the vendor improved the signature verification code in new
  versions of their firmware~\cite{florida_supplement} although
  existing machines that are not updated remain vulnerable, of
  course.}

\section*{Acknowledgments}

This research was supported by NSF grant CNS-0524252 and by a grant
from the Florida Department of State.  We also thank Avi Rubin for his
comments and support.

\bibliographystyle{abbrv}
\bibliography{rsa_padding_attack}

\end{document}